\documentclass[letter,twocolumn]{jpsj2} 
\title{Theory of ``Charge" Measured by the Shot Noise Experiments\\
in the Fractional Quantum Hall States}

\author{Daijiro \textsc{Yoshioka}}

\inst{Department of Basic Science, the University of Tokyo\\ 3-8-1 Komaba, Meguro-ku, Tokyo 153-8902\\
}

\abst{Shot noise at filling factor $\nu=2/5$ is investigated theoretically.
It is argued that the ``charge" $e^*$ measured by the noise at zero temperature is not the quasiparticle charge but simply the filling factor times the electron charge $e$, namely $e^*=2e/5$.
At higher temperature quasiparticles with charge $e^*=\pm e/5$ begin to contribute to the backscattering, and the shot noise gives charge $e^*=e/5$.
This theory explains recent experiment by Chung et al. [Phys. Rev. Lett. \textbf{91} (2003) 216804.]
}

\kword{fractional quantum Hall effect, quasiparticle charge, shot noise}

\def\runauthor{Daijiro {\sc Yoshioka}}

\begin{document}
\maketitle


Two-dimensional electrons in a strong magnetic field $B$ shows fractional quantum Hall effect at low temperature, when the filling factor of the lowest Landau level is in the vicinity of $\nu \equiv nh/eB = p/q$.\cite{tsui,dy}
Here $n$ is the electron density, $h$ is the Planck constant, $e$ is the charge of the electron, $p$ and $q$ are mutually prime integers.
The fractional quantum Hall state is characterized as an incompressible liquid state.
The charged excitation from this state, the quasiparticle, has been predicted to have charge $e^*=\pm e/q$, where the $\pm$ depends on whether the quasiparticle is a quasielectron or a quasihole.\cite{laughlin,haldane}

There have been several attempts for direct experimental observation of this charge.
In one of the experiments quantum antidot was used to measure the charge.\cite{goldman}
However, this measurement may not be direct, and it has been argued that different interpretation is possible.\cite{franklin}
Shot noise experiment, which we consider theoretically in this paper, has been considered to give more direct measurement of the quasiparticle charge.\cite{r3,r31,r4}
In the experiments, a constriction is placed in the two-dimensional systems, and the backscattered current created at the constriction is observed.
It is considered that the backscattered current consists of dilute flow of the quasiparticles, so the strength of the shot noise is proportional to the charge of the quasiparticle.\cite{r10,r7}

\begin{figure}[tb]
\begin{center}
\includegraphics[width=7cm]{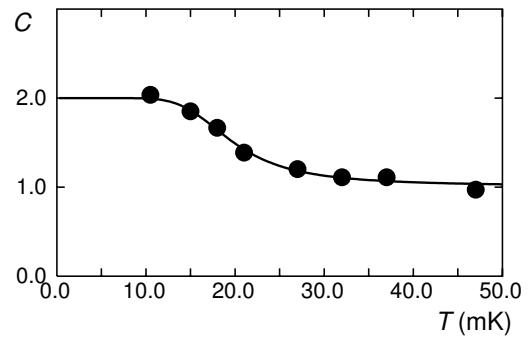}
\end{center}
\caption{Temperature dependence of the charge measured by the shot noise experiment.\cite{YCC}
The charge is given by $e^*=(C/5)e$.
The filled circles are the experimental data.
The solid line is the result of the present theory.}
\label{f1}
\end{figure}

Actually, quasiparticle charge $e/3$ at $\nu=1/3$,\cite{r3,r31} and $e/5$ at $\nu=2/5$\cite{r4} has been observed at relatively high temperature, $T \simeq 100$mK.
Especially the latter experiment is remarkable, because it in done in a situation where the charge is different from $\nu e$ nor the conductance times $e$.
However, a recent experiment at lower temperature shows something different.\cite{YCC}
The shot noise at $T > 40$mK gives a charge of $e^*=e/5$ at $\nu=2/5$, but the charge deduced from the noise gradually increase as the temperature is reduced.
At the lowest temperature $T=9$mK, the charge becomes $2e/5$ as shown in Fig.~\ref{f1}.
Similar increase of the deduced charge was observed also at $\nu=3/7$.

In the present letter we clarify the reason for such increase of the deduced charge.
We argue that at $T=0$ what is measured is not charge of the quasiparticle, but
 the filling factor of the fractional quantum Hall state.
The quasiparticle tunneling begins to be effective at higher temperature, and the measured charge approaches that of the quasiparticles.
Based on our theory we reproduce the temperature dependence in the experiment.
It is evident that we cannot obtain correct result, if we consider only quasiparticles.
In the present theory we consider the shot noise from various standpoints, which are electron picture, quasiparticle picture, and composite fermion picture.
Electrons are real, but the quasiparticles and composite fermions are objects introduced for convenient description of the phenomena.
This distinction is important for the construction of our theory.

\begin{figure}[tb]
\begin{center}
\includegraphics[width=7cm]{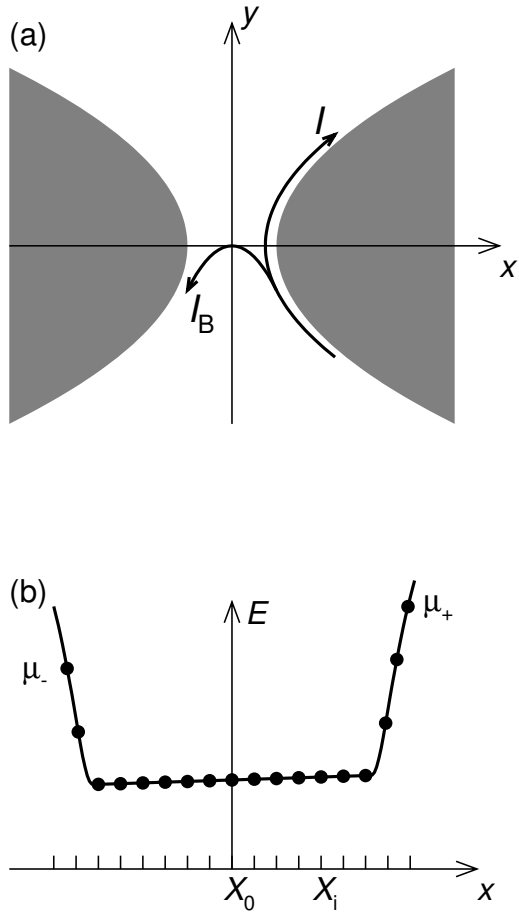}
\end{center}
\caption{(a) Geometry of the typical experiment.  Part of the current in the $y$-direction is backscattered at the constriction as $I_{\rm B}$.
(b) $x$ dependence of the Landau level at $y=0$.  The chemical potential at the right edge is $\mu_+$ and that at the left edge is $\mu_-$.
The single-electron states are labeled by the center coordinates $X_i$.}
\label{f2}
\end{figure}

Shot noise is measured in a geometry like that shown in Fig.~\ref{f2}(a).
A constriction to the two-dimensional plane is placed at the origin, and the current is flown in the $y$-direction.
Part of the current is backscattered at the constriction as $I_{\rm B}$, and the fluctuation in $I_{\rm B}$ is measured.
To understand the shot noise at low temperature, let us consider non-interacting 2-d electrons at $\nu=1$ quantum Hall state at first.
The single-electron states at the cross-section of the 2-d plane at $y=0$ can be specified by center coordinates of the wave function in the $x$-direction $X_i$, $i=0, \pm 1, \pm 2, \cdots$.
The spin-polarized electron occupies these $X_i$'s in the lowest Landau level up to the chemical potential $\mu_{\pm}$ as shown in Fig.~\ref{f2}(b).
The chemical potential at the right edge $\mu_+$ is higher than that of the left edge $\mu_-$.
The difference of the chemical potential gives the current in the $y$-direction, $I=(e/h)(\mu_+-\mu_-)$.
The backscattered current is created when electrons at the right edge are scattered to the left edge.\cite{note1}
For wide constriction, this scattering is rare, so the backscattered current is a dilute flow of electrons occupying just above the lower chemical potential $\mu_-$.
Since the electrons in dilute backscattered flow are not correlated, the shot noise is given by the classical formula,\cite{schottky}
\begin{equation}
S = 2 e I_{\rm B}\,.
\label{eq:1}
\end{equation}

Next we consider the case of the fractional quantum Hall state at $\nu=1/3$.
Also in this case the electrons occupy the single-electron states up to $\mu_{\pm}$.
However, due to the strong interaction between electrons, occupation probability of each state below $\mu_{\pm}$ is 1/3.
There are several ways to understand the shot noise in this case.
In the quasielectron picture, one can consider that each single-electron state is fully occupied by Laughlin quasielectrons of charge $e^*=e/3$ up tp $\mu_{\pm}$.
Similarly to the integer quantum Hall case, the elementary process, where backscattered current is created, is the scattering of the quasielectrons across the 2-d plane at the constriction.
Therefore, the shot noise is given by
\begin{equation}
S = 2 e^* I_{\rm B} = \frac{2}{3} e I_{\rm B} \,.
\label{eq:2}
\end{equation}
On the other hand, in the electron picture, the elementary process of quasielectron scattering is understood as translation of whole electron system to the adjacent center coordinates, namely electron at $X_i$ moving to $X_{i-1}$ at every site.
This translation moves charge $e/3$ from the right edge to the left edge, so the shot noise is also given by eq.(\ref{eq:2}).
Finally, we can also consider the shot noise by composite fermion picture.\cite{jain,r2}
In this picture, the effective magnetic field is reduced to 1/3, so the spacing between the center coordinates are expanded by factor three.
The number of center coordinates between the right and left edges are reduced by factor three.
These states are fully occupied by charge $e$ composite fermions.
Obviously, the elementary process for the backscattering is not a single composite fermion scattering from one side of the edge to the other.
Such a process is not the same as those by the quasielectron picture and the electron picture.
To describe the same process in the composite fermion picture we need to remember that there is a freedom to place the center coordinate for the composite fermions.
Namely, in one choice consecutive three electron center coordinates, $X_i$, $X_{i+1}$ and $X_{i+2}$, will be combined into a composite fermion center coordinates.
However, it is also possible to combine $X_{i-1}$, $X_{i}$ and $X_{i+1}$.
Therefore, for the composite fermion case it is possible to translate the whole composite fermions by a distance equal to the spacing between electron center coordinates, $\Delta X \equiv X_{i} - X_{i-1}$.
This translation moves charge $e/3$ from one edge to the other, and this is the same elementary process for the backscattered current as the other pictures.

After these preparations, we are now ready to understand the shot noise at $\nu=2/5$ at $T=0$.
In this case the single-electron states below the chemical potential are uniformly occupied with probability 2/5.
We first consider by the electron picture, which should be always valid.
The elementary process in this picture is the translation of the whole system by one step, $\Delta X$.
The charge transferred from right edge to the left edge is $2e/5$.
If we forget the confining potential, the one ground state tunnels into another ground state in this process; no excitation is involved.
This process is what determines the "charge" involved in the shot noise formula at $T=0$, so it is given by
\begin{equation}
S = 2\frac{2}{5} e I_{\rm B} \,.
\label{eq:3}
\end{equation}
In this description no quasielectron is involved, and what appears in the coefficient of the shot noise is just the average filling of the single-electron state, or the filling factor of the Landau level.
We cannot understand this process as a tunneling of a quasielectron whose charge is $e/5$.

\begin{figure}[tb]
\begin{center}
\includegraphics[width=7cm]{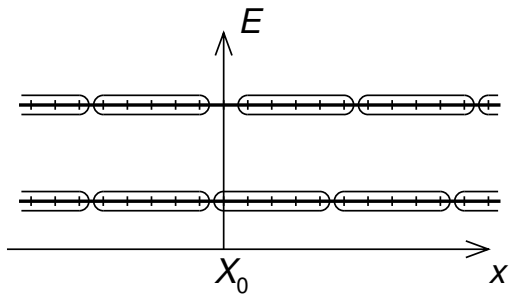}
\end{center}
\caption{Quasiparticles in the bulk of the $\nu=2/5$ fractional quantum Hall state.
Lowest two composite fermion Landau levels are shown, which are occupied by the composite fermions.
Short vertical bars on the Landau levels are the positions of the center coordinates for electrons.
One composite fermion state is related to five single-electron states as shown by closed curves.
If we shift the assignment of the composite fermion states to those of electrons at $x>X_0$ in one of the Landau levels, a quasihole of charge $e^*=e/5$ is created there.
}
\label{f3}
\end{figure}

Now let us consider the role of the quasiparticle and how the charge of the quasiparticle $e/5$ enters into the shot noise at higher temperature.
The quasiparticles at $\nu=2/5$ are best understood by the composite fermion picture.
Replacing the electrons with composite fermions that has two flux quanta attached in the opposite direction to the external field, we obtain effective field one-fifth of the original one, so the number of the center coordinates are reduced by five.
The composite fermions with charge $e$ occupy the lowest two Landau levels in the bulk.
To create a quasiparticle we use the freedom of the composite fermion center coordinates relative to the electron center coordinates.
Namely, we can change the selection of five electron states from which one CF states are composed as shown in Fig.~\ref{f3}.
In this figure a quasihole is created at $x=X_0$ in the upper Landau level.
Creation of the quasielectron is done similarly.
Now when all the composite fermions in the upper Landau level are translated by a distance $\Delta X$, quasiparticle is not created in the bulk.
Instead, charge $e^*=e/5$ is transferred from the right edge to the left edge.
This process can be considered as a scattering of a quasiparticle of charge $e^*$ from one edge to the other.
This process occurs at higher temperature.
However, we do not think this process effective at lower temperature.
The reason is that this process is not a transition between two ground states.
The configurations of the composite fermions in the upper and lower Landau levels are relatively shifted if this process occurs, and it should cost finite energy.
Thus the ``charge" at $T=0$ is $(2/5)e$.

Now we consider the temperature dependence of the ``charge" measured by the shot noise experiment.
Since we focus on the ``charge", and not on the temperature dependence of the noise itself, we neglect the thermal noise.
We take into account the effect of temperature through the frequency of the quasielectron tunneling, which we assume to be thermally activated with activation energy $E$.
Namely, we assume that the frequency of the ground state tunneling, in which charge $2e/5$ is transferred to be $n_1$, and that of the quasielectron tunneling to be $n_2\exp(-\beta E)$, where $\beta=1/k_{\rm B}T$ is inverse temperature.
Then the backscattered current is given by
\begin{equation}
I_{\rm B} = i_1 + i_2 = q_1n_1 + q_2n_2\exp(-\beta E) ,
\label{eq:4}
\end{equation}
where $q_1=2e/5$ and $q_2=e/5$.
At weak constriction each tunneling occurs independently.
Then the shot noise is given by a summation of noise from each processes.
\begin{equation}
S = 2q_1 i_1 + 2 q_2 i_2 = 2 q_1^2 n_1 + 2 q_2^2 n_2\exp(-\beta E).
\label{eq:5}
\end{equation}
In the experiment the ratio $S/2I_{\rm B}$ is measured as ``charge", thus it is expressed as
\begin{eqnarray}
\frac{S}{2I_{\rm B}}
&=&
\frac{n_1q_1^2 + n_2\exp(-\beta E) q_2^2}{n_1q_1 + n_2\exp(-\beta E) q_2}\nonumber\\
&=&
\frac{4n_1 + n_2\exp(-\beta E)}{2n_1 + n_2\exp(-\beta E)}\, \left(\frac{1}{5}\right)e.
\label{eq:6}
\end{eqnarray}
In this equation we have two parameters $n_2/n_1$ and $E$ to be determined by the experimental data.
By least-square-fitting of the data shown in Fig.~\ref{f1}, we obtain $n_2/n_1=565$ and $E/k_{\rm B}=112$mK.
Using these values we plot eq.(\ref{eq:6}) as a solid curve in Fig.~\ref{f1}.
The agreement is satisfactory.


At present stage, we cannot give theoretical estimate of the parameters.
The ratio $n_2/n_1$ will be given by the tunneling probabilities of the ground state and the quasielectron states, namely the translation of the whole electron system and translation of the composite fermions in the higher Landau level, respectively.
It is natural that the matrix element of the former translation is exponentially smaller than that of the latter, so $n_2/n_1$ is large.
The activation energy $E$ is much smaller than the typical activation energy of the diagonal resistivity at $\nu=2/5$, which is of the order of 1K.\cite{morf}
If we consider naively that the quasielectron scattering is equivalent to quasielectron-quasihole pair excitation at the edges, the difference of the energy by an order of magnitude is quite strange.
However, one should remember that the excitation energy at the edge is gapless.
Thus, if the two Landau levels of the composite fermions have little correlation the energy can be small.
Theoretical estimate of this energy taking into account various experimental details such as confining potential around the constriction, thickness, impurity, etc. is not easy, so we leave it as a future task.


The present theory is applicable to other fractional quantum Hall states, such as that at $\nu=3/7$.
The experiment at $\nu=3/7$ also has shown increase of the ``charge" at lower temperature.\cite{YCC}
Except for the fundamental quantum Hall state at $\nu=1/q$, the picture of quasiparticle tunneling is not appropriate at $T=0$.
The ``charge" determined by the shot noise experiment at $T=0$ is just filling factor times the electron charge, and not the charge of the quasiparticles.
When the composite fermion occupies plural Landau levels, moving a quasiparticle in one of the Landau levels costs finite energy, so this process is not effective at low temperature.

As a test for the present theory we suggest to do experiments at higher Hall voltages.
In such a case, it will be possible that the energy gain by the tunneling $(e^*/e)(\mu_+ - \mu_-)$ is larger than the activation energy $E$.
Then we expect increase of the backscattered current, and decrease of the ``charge" measured by the shot noise.
The increase of the backscattered current observed in ref.\citen{YCC}, Fig.2a may be related to this possibility.

\section*{Acknowledgment}
The author thanks Prof. Yunchul Chung and Prof. Moty Heiblum for telling me details of the experiment and providing me raw experimental data.
He also appreciates hospitality of the Aspen Center for Physics where part of this work was done.
This work is supported by Grant-in Aid No. 14540294 from JSPS.



\begin{thebibliography}{99}
\bibitem{note1} The scattering of one electron between edges is the same as translation of the whole electron system by $\Delta X \equiv X_{i+1} - X_{i}$.
In this paper we call this process as scattering for simplicity.
\end{thebibliography}
\end{document}